\documentclass[aps,prb,twocolumn,10pt,floatfix]{revtex4-2}
\usepackage{amssymb}
\usepackage{amsmath}
\usepackage{lmodern}
\usepackage{graphicx}
\usepackage{cases} 
\usepackage[mathcal]{eucal}
\usepackage[american]{circuitikz}
\usepackage[hidelinks]{hyperref}
\newcommand{\scaledint}
{\mathop{\scalebox{1.2}{$
\displaystyle\int$}}\limits}

\begin{document}

\title{Shocks in a Lossy Nonlinear Transmission Line: The Inverse-Design Approach}

\author{Eugene Kogan}
\email{Eugene.Kogan@biu.ac.il}
\affiliation{Department of Physics, Bar-Ilan University, Ramat Gan 5290002, Israel}

\begin{abstract}
We formulate a traveling-wave equation describing traveling fronts in
a lossy transmission line with a nonlinear voltage--charge relation.
Within the inverse-design approach, we start from a prescribed front
profile and determine exactly the voltage--charge relation supporting it,
together with the associated squared local sound speed, whose finiteness
and positivity determine whether the front is physically admissible.
We specifically consider fronts whose profiles belong to the five-parameter asymmetric power-logistic family.
For such fronts, the voltage--charge relation and the squared local sound speed can be expressed in terms
of the lower incomplete beta function.
All physically admissible fronts of this form are shocks rather than kinks.
We formulate a variational method for approximately solving the traveling-wave equation
when the voltage--charge relation is
prescribed. For weak shocks and a cubic voltage-charge relation, the approximate solution of the direct problem coincides with the asymptotic
expansion of the exact solution of the inverse problem at next-to-leading order; under a certain condition, also at the next order.
Finally, a local spectral analysis of the original partial differential
equations shows that short-wavelength perturbations are damped when the
squared front speed and the squared local sound speed are positive. A
long-wavelength characteristic analysis
distinguishes between compressive shocks and noncompressive kinks.
\end{abstract}

\date{\today}

\maketitle

\section{Introduction}

Nonlinear electrical transmission lines serve both as useful circuit
elements and as experimentally accessible models of nonlinear-wave
propagation. A voltage-dependent capacitance or current-dependent inductance
makes the wave speed depend on the wave amplitude,
leading to wave steepening, harmonic generation, solitary waves, and
shocks. In an early analysis, Landauer showed that capacitive nonlinearity can
transform a periodic signal into an electromagnetic shock and emphasized
the role of dissipation in determining the subsequent shock
evolution~\cite{landauer}. Electrical networks were later used as controlled
lattices for observing soliton propagation and
interaction~\cite{hirota}. Soliton-like signals in nonlinear electrical
transmission lines have also been constructed, and their linear
stability has been analyzed within a modified complex Ginzburg--Landau
description~\cite{kengne}. Related multiscale reductions of composite
right- and left-handed nonlinear transmission lines predict coupled
backward- and forward-propagating vector solitons~\cite{veldes}.
Experiments on nonlinear left-handed transmission lines have also
demonstrated multistability, period doubling, and the onset of chaotic
dynamics at higher input powers~\cite{powell}.
Stationary and traveling localized modes have been observed
experimentally and reproduced numerically in driven two-dimensional
nonlinear electrical lattices~\cite{english}. More generally, the
existence, stability, and velocity-dependent dynamics of moving
discrete breathers have been investigated in nonlinear
$\beta$-Fermi--Pasta--Ulam lattices~\cite{duran}.

Nonlinear transmission lines also have direct technological applications. Varactor-loaded nonlinear transmission lines compress electrical
wave fronts and generate picosecond pulses~\cite{madden}. Nonlinear $LC$ transmission lines
with cubic capacitive
nonlinearity have also been proposed as analog-gravity systems, in
which a voltage soliton creates an effective event horizon for
electromagnetic waves~\cite{katayama}. On the mathematical side,
solutions of continuum models with
voltage-dependent capacitance, resistance, and leakage conductance have
been shown to develop shock discontinuities from sufficiently steep
initial data~\cite{bloom}. Thus, nonlinearity supplies the steepening
mechanism, while dissipation regularizes the discontinuity into a
traveling front of finite width.

Factorization approaches
provide an analytical route to solving nonlinear second-order
differential equations with polynomial
nonlinearities~\cite{gonzalez}. Related exact-solution and integrability
techniques have been applied to the Duffing--van der Pol
equation~\cite{kudryashov}. Kaliappan obtained exact traveling-wave solutions
of a nonlinear reaction--diffusion equation~\cite{kaliappan}.

Inverse constructions have also been developed for discrete nonlinear
systems. Flach, Zolotaryuk, and Kladko prescribed a traveling kink or
pulse and its velocity and then determined lattice equations supporting
that solution~\cite{flach-zolotaryuk}. This approach was subsequently
discussed in the broader context of generic and nongeneric
differential--difference equations, Peierls--Nabarro barriers, and
depinning transitions~\cite{kevrekidis-zolotaryuk}. In contrast, the
present work applies inverse design to the continuum equations of a
lossy nonlinear transmission line and determines the constitutive
voltage--charge relation required to support a prescribed front.

In our previous work, we derived a traveling-wave equation for lossy
transmission lines with nonlinear capacitance and obtained an approximate
traveling-front solution~\cite{kogan6,kogan7}. More recently, in
Ref.~\cite{kogan8}, we developed an exact inverse-design approach for
traveling fronts with a prescribed hyperbolic-tangent profile. Rather than
specifying the nonlinear voltage--charge relation and subsequently solving
the traveling-wave equation, we prescribed the front profile and its
propagation speed and determined the constitutive relation required to
support that profile. The dimensionless squared local sound speed was
obtained exactly in terms of the lower incomplete beta function. We
established its monotonic increase across the front and derived the
corresponding physical admissibility conditions, showing that all
physically admissible hyperbolic-tangent fronts are shocks rather than
kinks. We also considered the direct problem for a prescribed cubic
voltage--charge relation. An approximate solution of the traveling-wave equation was constructed
by minimizing its mean-square residual, and was shown to agree with the exact inverse construction in the broad-shock limit. Finally, we investigated
the local spectral properties of the fronts and related the distinction
between shocks and kinks to their long-wavelength characteristic structure.

The present work substantially generalizes the inverse-design construction
of Ref.~\cite{kogan8}. Instead of restricting the analysis to the symmetric
hyperbolic-tangent profile, we consider a five-parameter family of
asymmetric traveling fronts, specified through the charge
and traveling-wave coordinate. This generalization permits different
asymptotic behaviors at the two edges of the front and includes the
previously studied hyperbolic-tangent profile as a particular case. We also
derive the corresponding voltage--charge relation and squared local sound
speed, which can again be expressed in terms of the lower incomplete beta
function. Thus, the role of this special function is extended from the
symmetric hyperbolic-tangent profile to a broader family of traveling
fronts.

Another difference concerns the treatment of the front boundaries.
Whereas Ref.~\cite{kogan8} derived explicit limiting sound speeds for
hyperbolic-tangent fronts and considered more general exponential endpoint
asymptotics, the present analysis formulates the endpoint behavior for
general prescribed profiles. This allows the admissibility
conditions to be investigated separately at the two asymptotic states and
provides a framework for examining asymmetric fronts. We also derive
asymptotic expansions of the voltage--charge relation near the shock edges.

The treatment of the direct problem is modified as well. In
Ref.~\cite{kogan8}, the parameters of an approximate hyperbolic-tangent
solution were determined by minimizing the discrepancy between the
prescribed squared sound speed and that inferred from the trial profile.
In the present work, we instead minimize the residual of the original
traveling-front equation. This change makes the approximation criterion
directly dependent on the equation of motion and permits a more general
choice of trial profiles.  
We also show that, in the weak-shock limit, the approximate solution of the direct problem reproduces the leading terms of the asymptotic expansion of the exact inverse-problem solution. The local spectral analysis is
retained to relate the admissibility and classification of the resulting
fronts to the propagation of small disturbances.

\section{The Lossy Transmission Line: Kirchhoff's Laws}
\label{circ}

The transmission line studied here, shown in Fig.~\ref{r5}, consists of
identical inductive branches, each shunted by a resistor $R_L$. Each node is
connected to ground through a nonlinear capacitor in series with a resistor
$R_C$.
\begin{figure}[!t]
\centering
\resizebox{\columnwidth}{!}{%
\begin{circuitikz}[
  line width=0.8pt,
  every node/.style={font=\large}
]
\draw (0,3) to[L,l=$L$,v_=$\mathcal{V}_{n}$,i>^=$I_n$] (4,3)
            to[L,l=$L$,v_=$\mathcal{V}_{n+1}$,i>^=$I_{n+1}$] (8,3);
  \draw (0,3) -- (0,4.35) to[R,l=$R_L$] (4,4.35) -- (4,3);
  \draw (4,3) -- (4,4.35) to[R,l=$R_L$] (8,4.35) -- (8,3);

  \draw (0,3) to[vC,l_=$Q_{n-1}$,v^>=$V_{n-1}$] (0,1.45)
              to[R,l_=$R_C$] (0,0);
  \draw (4,3) to[vC,l_=$Q_n$,v^>=$V_n$] (4,1.45)
              to[R,l_=$R_C$] (4,0);
  \draw (8,3) to[vC,l_=$Q_{n+1}$,v^>=$V_{n+1}$] (8,1.45)
              to[R,l_=$R_C$] (8,0);

  \draw (-0.35,0) -- (8.35,0);
  \draw (4,0) node[ground] {};
  \fill (0,3) circle (1.6pt);
  \fill (4,3) circle (1.6pt);
  \fill (8,3) circle (1.6pt);
\end{circuitikz}%
}
\caption{Unit cells of the lossy nonlinear transmission line.}
\label{r5}
\end{figure}
We denote the charge on and the voltage across the $n$th nonlinear capacitor
by $Q_n$ and $V_n$, respectively. The current through the $n$th inductive
branch is denoted by $I_n$, and the voltage across it by
$\mathcal V_n$. Kirchhoff's laws then give
\begin{equation}
\label{lb}
\begin{aligned}
\frac{dQ_n(t)}{dt}&=I_n(t)-I_{n+1}(t)+\frac{1}{R_L}
\Bigl[\mathcal V_n(t)-\mathcal V_{n+1}(t)\Bigr],
\\[3mm]
\mathcal V_{n+1}(t)&=V_n(t)-V_{n+1}(t)+R_C\frac{d}{dt}\Bigl[Q_n(t)-Q_{n+1}(t)\Bigr].
\end{aligned}
\end{equation}
We restrict the analysis that follows to transmission lines
with a linear inductance
\begin{equation}
\mathcal{V}_n=L\frac{dI_n}{dt}, 
\end{equation}
where $L$ is a constant, and a nonlinear
voltage--charge relation $V_n=V(Q_n)$. The parameters $L$, $R_L$, and $R_C$ are considered prescribed throughout.

In the continuum approximation, we replace the discrete index $n$ by the
continuous coordinate $z$, measured in units of the transmission-line period,
and approximate the finite differences on the right-hand sides by first
derivatives with respect to $z$. In this approximation, Eq.~(\ref{lb}) takes the form
\begin{equation}
\label{c2b}
\begin{aligned}
&\partial_t Q(z,t)=-\partial_z I(z,t)-\frac{L}{R_L}\partial^2_{zt}I(z,t),\\[3mm]
&L\partial_t I(z,t)=-\partial_z V\bigl(Q(z,t)\bigr)-R_C\partial^2_{zt}Q(z,t).
\end{aligned}
\end{equation}

We consider traveling waves for which all quantities
depend on the single dimensionless variable
\begin{equation}
\theta=\frac{R_L(t-z/U)}{L},
\end{equation} 
where $U>0$ is the wave speed (for the sake of definiteness, everywhere in this paper we consider the waves propagating from left to right). 
For such waves, Eq.~(\ref{c2b}) reduces to the system of ordinary
differential equations
\begin{equation}
\label{low}
\begin{aligned}
\Bigl[\widetilde I^{\displaystyle\prime}(\theta)+\widetilde I(\theta)-Q(\theta)\Bigr]^{\displaystyle\prime}&=0,
\\[2mm]
\Bigl[Q^{\displaystyle\prime}(\theta)+\widetilde V\bigl(Q(\theta)\bigr)
-W\widetilde I(\theta)\Bigr]^{\displaystyle\prime}&=0,
\end{aligned}
\end{equation}
where $\widetilde V\equiv LV/(R_CR_L)$, $\widetilde I=I/U$ and $W\equiv L^2U^2/(R_CR_L)$ is the dimensionless squared wave speed. Throughout, a prime denotes differentiation with respect to the displayed argument.

Integrating Eq.~(\ref{low}) once, we obtain the system of two first-order ordinary
differential equations:
\begin{align}
\widetilde I^{\displaystyle\prime}(\theta)+\widetilde I(\theta)-Q(\theta)&=0,\label{A}
\\[2mm]
Q^{\displaystyle\prime}(\theta)
+\widetilde V\bigl(Q(\theta)\bigr)-W\widetilde I(\theta)&=0,\label{B}
\end{align}
where $\widetilde I(\theta)$ and $\widetilde V(Q)$ are defined up to arbitrary
additive constants.

\section{Traveling-Wave Equation}

\subsection{Li\'enard Equation}

Solving Eq.~(\ref{B}) for $\widetilde I(\theta)$ and substituting
the result into Eq.~(\ref{A}), we obtain the closed traveling-wave equation
for $Q(\theta)$ of Li\'enard type:
\begin{equation}
\label{199}
Q^{\displaystyle\prime\prime}(\theta)+\Bigl[w\bigl(Q(\theta)\bigr)
+1\Bigr]Q^{\displaystyle\prime}(\theta)
=F\bigl(Q(\theta)\bigr),
\end{equation}
where
\begin{align}
F(Q)&\equiv WQ-\widetilde V(Q)-\widetilde V_0,\label{foy}\\[2mm]
w(Q)&\equiv\widetilde V^{\displaystyle\prime}(Q),\label{foy2}
\end{align}
and $\widetilde V_0$ is the integration constant (a linear combination of the integration constants entering in Eqs.~\eqref{A} and~\eqref{B}).
Note that 
\begin{equation}
w(Q)=\frac{L^2}{R_C R_L}u^2(Q),
\end{equation}
where $u(Q)$ is the speed of small-amplitude,
long-wavelength (sound) waves on a homogeneous background with charge $Q$~\cite{kogan6}.
The dimensionless squared local sound speed $w(Q)$ 
will play the central role in our analysis. 

\subsection{Shocks and Kinks}
\label{ksp}

We impose the traveling front boundary conditions
\begin{equation}
\label{bc}
\lim_{\theta\to-\infty}Q(\theta)=Q_1,\qquad\lim_{\theta\to\infty}Q(\theta)=Q_2.
\end{equation}
Throughout this paper $Q_1$ and $Q_2$ are considered to be prescribed.
It follows from the boundary conditions that
\begin{equation}
\label{norma}
\begin{cases}
\widetilde V(Q_1)+\widetilde V_0=Q_1W,\\[2mm]
\widetilde V(Q_2)+\widetilde V_0=Q_2W.
\end{cases}
\end{equation}
Eq.~\eqref{norma} fixes the squared front speed
\begin{equation}
\label{norma1}
W=\frac{\widetilde V(Q_2)-\widetilde V(Q_1)}{Q_2-Q_1}
\end{equation}
and the additive constant in $\widetilde V$
\begin{equation}
\label{norma12}
\widetilde V_0=\frac{Q_1\widetilde V(Q_2)-Q_2\widetilde V(Q_1)}{Q_2-Q_1}.
\end{equation}
Integrating Eq.~(\ref{foy2}) with respect to $Q$ from $Q_1$ to $Q_2$
and using Eq.~(\ref{norma}), we obtain
\begin{equation}
\label{wonder}
W=\frac{1}{Q_2-Q_1}\int\limits_{Q_1}^{Q_2} w(Q)\,dQ,
\end{equation}
that is, the squared front speed is equal to the squared sound speed, averaged uniformly over the charge interval spanned by the front.

Equation~(\ref{199}) with the boundary conditions~\eqref{norma} admits an obvious geometrical-mechanical
interpretation. We start from the function $\widetilde V(Q)$.
Then we construct a straight line joining the points $\bigl(Q_1,\widetilde V(Q_1)\bigr)$ and $\bigl(Q_2,\widetilde V(Q_2)\bigr)$. The function $F(Q)$ is the elevation of the straight line above the function $\widetilde V(Q)$. 
Equation~(\ref{199}) describes
the motion of a fictitious particle of unit mass with coordinate $Q$ and
time $\theta$ in the field of force $F(Q)$.
The motion is subject to viscous damping,
with the coordinate-dependent effective damping coefficient $1+\widetilde V^{\displaystyle\prime}(Q)$.

The distinction between kinks and shocks can be formulated in several ways.
Let us first use the language of mechanics.
The particle starts its motion at $Q=Q_1$, which is a local maximum of the potential
$\Pi(Q)$ defined by the equation
\begin{equation}
\label{pipi}
\Pi^{\displaystyle\prime}(Q)=\widetilde V(Q)+\widetilde V_0-WQ,
\end{equation}
and ends its motion at the point
$Q=Q_2$, which is a point of maximum for a kink and of minimum for a shock.

In the language of hydrodynamics, the distinction is formulated in terms of the speeds. 
Combining Eqs.~(\ref{foy2}) and~(\ref{pipi}), we obtain
\begin{equation}
w(Q)=W+\Pi^{\displaystyle\prime\prime}(Q).
\end{equation}
We see that the traveling front is supersonic relative to the 
asymptotic state ahead of it. 
Relative to the asymptotic
state behind it, the front is also supersonic for a kink but subsonic
for a shock~\cite{whitham,kogan6}.
\begin{equation}
\begin{cases}
&w(Q_1)<W,\\[2mm]
&w(Q_2)\begin{cases}
<W, \qquad\text{(kink)}, \\[2mm]
>W, \qquad\text{(shock)}.
\end{cases}
\end{cases}
\end{equation} 

Finally, in the language of geometry, assuming for the sake of definiteness that $Q_2>Q_1$,
the straight line we mentioned previously, crosses the graph of the function $\widetilde V(Q)$ at the point $Q=Q_1$ from below, when we move from left to right. It crosses the same graph also from below for a kink and from above for a shock.
Additional details of the kink/shock dichotomy are given in
Sec.~\ref{stab}.

\subsection{Weak Shocks}

Consider Eq.~\eqref{199} in the case of weak shocks: $\Delta Q\to 0$
for fixed $V(Q)$.
In the leading order,
\begin{equation}
\begin{split}
F(Q)&=\frac{1}{2}w^{\displaystyle\prime}
(\overline Q)\bigl(Q-Q_1\bigr)\bigl(Q_2-Q\bigr)=O\bigl((\Delta Q)^2\bigr),
\end{split}
\end{equation}
where
\begin{equation}
\overline Q\equiv \frac{Q_2+Q_1}{2},\qquad \Delta Q\equiv \frac{Q_2-Q_1}{2}.
\end{equation}
We see that the weak shocks are the broad shocks with
\begin{equation}
Q^{\displaystyle\prime}(\theta)=O\bigl((\Delta Q)^2\bigr).
\end{equation}
Hence
\begin{equation}
Q^{\displaystyle\prime\prime}(\theta)=O\bigl((\Delta Q)^3\bigr)
\end{equation}
and can be discarded in the leading order. 
Hence, in the leading order Eq.~\eqref{199} takes the form
\begin{equation}
\label{1977}
(W+1)Q^{\displaystyle\prime}(\theta)=\frac{1}{2}w^{\displaystyle\prime}
(\overline Q)\bigl(Q-Q_1\bigr)\bigl(Q_2-Q\bigr).
\end{equation}
Eq.~\eqref{1977} can be easily solved, and we obtain the hyperbolic-tangent profile
\begin{equation}
\label{197}
Q(\theta)=\overline Q+\Delta Q\tanh\left(\frac{\theta}{2m}\right),
\end{equation}
where
\begin{equation}
\label{1976}
m=\frac{W+1}{w^{\displaystyle\prime}(\overline Q)\,\Delta Q}.
\end{equation}
In Sec.~\ref{appro1} we will see how this approximation can be improved using the variational principle, to be presented below.

\subsection{The Variational Principle for the Li\'enard Equation}

Assuming that the solution of Eq.~\eqref{199} subject to the boundary
conditions~\eqref{bc} is strictly monotone, with
$Q^{\displaystyle\prime}(\theta)\neq0$ for finite $\theta$, 
we can present Eq.~\eqref{199} as the equation for the function $\theta^{\displaystyle\prime}(Q)$
\begin{equation}
\label{18}
\mathcal R(Q)\equiv\left[\frac{1}{\theta^{\displaystyle\prime}(Q)}\right]^{\displaystyle\prime}
+\widetilde V^{\displaystyle\prime}(Q)+1+\bigl[\widetilde V(Q)+\widetilde V_0-WQ\bigr]
\theta^{\displaystyle\prime}(Q)=0
\end{equation}
with the boundary conditions
\begin{equation}
\label{bbcc}
\frac{1}{\theta^{\displaystyle\prime}(Q)}\to0,\qquad Q\to Q_1,Q_2.
\end{equation}
Solving the
differential equation \eqref{18} is equivalent to finding the function $\theta^{\displaystyle\prime}(Q)$,
for which the following functional vanishes
\begin{equation}
\label{acto}
{\mathcal F}\bigl[\theta^{\displaystyle\prime}(Q)\bigr]
=\frac{1}{Q_2-Q_1}\int\limits_{Q_1}^{Q_2}\mathcal R^2(Q)\,dQ,
\end{equation}
and the solution of the overdetermined ODE
\begin{equation}
\label{over}
Q^{\displaystyle\prime}(\theta)=\frac{1}{\theta^{\displaystyle\prime}_{\mathrm{min}}(Q)},
\end{equation}
with the boundary conditions~\eqref{bc} exists.

The variational reformulation of the Li\'enard equation suggests the
following approximation method. 
We choose a trial function $\theta^{\displaystyle\prime}_{\mathrm{tr}}(Q)$ satisfying the boundary conditions~\eqref{bbcc} and
depending on fitting parameters.
The functional ${\mathcal F}\bigl[\theta^{\displaystyle\prime}(Q)\bigr]$ then becomes a function ${\mathcal F}_{\mathrm tr}(\dots)$ of these parameters.
We choose the values of the parameters minimizing
the function and obtaining $\theta^{\displaystyle\prime}_{\mathrm{min}}(Q)$. Solving 
for $Q(\theta)$ Eq.~\eqref{over}, 
we obtain an approximate solution of the original problem.
In the framework of the variational principle, the squared front speed $W$ and the additive constant in $\widetilde V$ are treated initially as the free parameters. They are determined in the process of the minimization.
 We will see how it works in Sec.~\ref{appro1}.

\subsection{The Integro-Differential Equation}
\label{ksp22}

To obtain an integral formulation of the traveling-wave problem, it is
convenient to introduce the exponential time
$\Theta=\exp(\theta)$.
Since $\theta$ is defined up to an additive constant, $\Theta$ is
defined up to a positive multiplicative constant. In terms of this
variable, Eq.~\eqref{A} becomes (starting from here we include the integration constants into $\widetilde I(\Theta)$ and $\widetilde V(\Theta)$)
\begin{align}
\Theta\widetilde I^{\displaystyle\prime}(\Theta)+\widetilde I(\Theta)-Q(\Theta)&=0,
\label{AC}\\[3mm]
\Theta Q^{\displaystyle\prime}(\Theta)
+\widetilde V\bigl(Q(\Theta)\bigr)-W\widetilde I(\Theta)&=0, \label{BC}
\end{align}
Eq.~(\ref{AC}) can be integrated by quadrature,
\begin{equation}
\label{A2}
\widetilde I(\Theta)=\frac{1}{\Theta}\int\limits_0^\Theta
Q(\tau)\,d\tau,
\end{equation}
where the lower integration limit has been chosen so that
$\widetilde I(\Theta)$ remains finite as $\Theta\to0$.
Substituting Eq.~\eqref{A2} into Eq.~\eqref{BC}, we obtain the
integro-differential equation for $Q(\Theta)$:
\begin{equation}
\label{B20}
\widetilde V\bigl(Q(\Theta)\bigr)=W\widetilde I(\Theta)
-\Theta Q^{\displaystyle\prime}(\Theta),
\end{equation}
where $\widetilde I(\Theta)$ is given by Eq.~\eqref{A2}.
Unlike the differential equation~\eqref{199}, Eq.~\eqref{B20} itself implies the boundary conditions~\eqref{norma} (with $\widetilde V_0=0$), which can be checked
using L'Hôpital's rule.

\section{The Inverse-Design Approach}
\label{iver}

\subsection{The Specified Profile}

So far, we have treated $\widetilde V(Q)$ as prescribed.
In the framework of the inverse-design approach, the function
$\widetilde V(Q)$ is treated as unknown. Instead, we prescribe the front profile 
\begin{equation}
\label{hn}
Q=Q_{\mathrm{fr}}(\Theta),
\end{equation}
where $Q_{\mathrm{fr}}(\Theta)$ is smooth and strictly monotonic,
with
\begin{equation}
Q_{\mathrm{fr}}^{\displaystyle\prime}(\Theta)\neq0,\qquad \Theta>0.
\end{equation}
We also prescribe the squared front speed $W$. With this change
of viewpoint, Eq.~\eqref{B20} ceases to be an equation for
determining $Q(\Theta)$ and (together with Eq.~\eqref{hn}) becomes the exact voltage--charge relation $\widetilde V(Q)$ in parametric representation, which after integration by parts takes the form
\begin{equation}
\label{B21}
\widetilde V=WQ_{\mathrm{fr}}(\Theta)-\frac{W{\mathcal I}(\Theta)}{\Theta}-{\mathcal Q}(\Theta),
\end{equation}
where
\begin{align}
{\mathcal I}(\Theta)&\equiv \int\limits_0^{\Theta} 
\tau Q_{\mathrm{fr}}^{\displaystyle\prime}(\tau)\,d\tau,\label{balan1}\\[2mm]
{\mathcal Q}(\Theta)&\equiv\Theta Q_{\mathrm{fr}}^{\displaystyle\prime}(\Theta)
=Q_{\mathrm{fr}}^{\displaystyle\prime}(\theta).
\end{align}
Differentiating Eq.~\eqref{B21} with respect to $Q$, we obtain for the local squared sound speed $w=d\widetilde V/dQ$ (also in the parametric representation):
\begin{equation}
\label{B25}
w=\frac{W{\mathcal I}(\Theta)}
{\Theta{\mathcal Q}(\Theta)}
-\frac{\mathcal Q^{\displaystyle\prime}(\Theta)}
{Q^{\displaystyle\prime}_{\mathrm{fr}}(\Theta)}.
\end{equation}
Repeated integration by parts in the integral in Eq.~\eqref{balan1}
gives the asymptotic expansion~\cite{wong}:
\begin{equation}
\label{B21b}
{\mathcal I}(\Theta)\sim 
\sum_{p=1}^{\infty}\frac{(-1)^{p+1}\Theta^{p+1}}{(p+1)!}\,Q_{\mathrm{fr}}^{(p)}(\Theta).
\end{equation}

\subsection{The Endpoint Conditions and the Local Sound Speed}

Because of the boundary conditions~\eqref{bc}, ${\mathcal Q}(\Theta)$ goes to zero at both endpoints.
Assume that it goes to zero at both ends according to a power law
\begin{numcases}{}
{\mathcal Q}(\Theta)\propto\Theta^{1/\widetilde m_1},\qquad \Theta\to0,\quad\widetilde m_1>0,
\label{symptotic}
\\[2mm]
{\mathcal Q}(\Theta)\propto\Theta^{-1/\widetilde m_2},
\qquad \Theta\to\infty,\quad\widetilde m_2>0.\label{sympotic}
\end{numcases}
For $w(Q_2)$ to be finite, we need that the integral~\eqref{balan1} diverges as a power of $\Theta$ when $\Theta\to\infty$, which requires
\begin{equation}
\label{plus}
\widetilde m_2>1.
\end{equation}
Hence the numerator and the denominator of the first fraction in Eq.~\eqref{B25} go to zero when $\Theta\to 0$ and to $\infty$ when $\Theta\to\infty$, 
and to find both $w(Q_1)$ and $w(Q_2)$ we may use
L'Hôpital's rule, which gives
\begin{numcases}{}
\displaystyle w(Q_1)=\frac{\widetilde m_1W}{\widetilde m_1+1}-\frac{1}{\widetilde m_1},\label{pluh}\\[3mm]
\displaystyle w(Q_2)=\frac{\widetilde m_2W}{\widetilde m_2-1}+\frac{1}{\widetilde m_2}.\label{pluha}
\end{numcases}
Eq.~\eqref{pluha} gives $w(Q_2)>W$.
Hence every physically admissible front with the asymptotic behavior at the trailing edge given by Eq.~(\ref{sympotic}) is necessarily a shock.
The positivity of $w(Q_1)$ requires
\begin{equation}
\label{minus}
W>\frac{\widetilde m_1+1}{\widetilde m_1^2}.
\end{equation}

The inequality~(\ref{minus}) is the endpoint condition. 
The inequality
\begin{equation}
\label{soundd4}
w(Q)>0,\qquad Q\in
\bigl(\min\{Q_1,Q_2\},\max\{Q_1,Q_2\}\bigr)
\end{equation}
is the interior physical admissibility condition for the prescribed front profile.
It should be checked separately.

\subsection{Front Profile Parametric Representation}

More generally, we may prescribe the front profile in the
parametric form
\begin{align}
\Theta=\Theta_{\mathrm{fr}}(Y),\\[2mm]
Q=Q_{\mathrm{fr}}(Y),\label{parva}
\end{align}
where $Q_{\mathrm{fr}}$ and $\Theta_{\mathrm{fr}}$ are smooth, strictly
monotonic functions on an interval $(Y_1,Y_2)$, with $\Theta_{\mathrm{fr}}$
increasing, and
\begin{equation}
\label{bbc}
\begin{aligned}
&\lim_{Y\to Y_1^+}\Theta_{\mathrm{fr}}(Y)=0,\qquad &&\lim_{Y\to Y_1^+}
Q_{\mathrm{fr}}(Y)=Q_1,\\
&\lim_{Y\to Y_2^-}\Theta_{\mathrm{fr}}(Y)=\infty,\qquad &&\lim_{Y\to Y_2^-}
Q_{\mathrm{fr}}(Y)=Q_2.
\end{aligned}
\end{equation}
We assume, in addition, that
\begin{equation}
Q_{\mathrm{fr}}^{\displaystyle\prime}(Y)\neq0,
\qquad
\Theta_{\mathrm{fr}}^{\displaystyle\prime}(Y)>0,
\qquad Y\in(Y_1,Y_2).
\end{equation}

For a parametrically specified front profile, Eqs.~\eqref{B21} and~\eqref{B25} are modified to
\begin{align}
\widetilde V&=WQ_{\mathrm{fr}}(Y)-\frac{W{\mathcal I}(Y)}{\Theta_{\mathrm{fr}}(Y)}-{\mathcal Q}(Y),  \label{49}\\[4mm]
w&=\dfrac{W{\mathcal I}(Y)}{\Theta_{\mathrm{fr}}(Y){\mathcal Q}(Y)} 
-\frac{\mathcal Q^{\displaystyle\prime}(Y)}
{Q_{\mathrm{fr}}^{\displaystyle\prime}(Y)},    \label{48}
\end{align}
where
\begin{align}
{\mathcal I}(Y)&=
\int\limits_{Y_1}^{Y}\Theta_{\mathrm{fr}}(y)Q^{\displaystyle\prime}_{\mathrm{fr}}(y)dy,\label{8h9}\\[3mm]
{\mathcal Q}(Y)&=\frac{\Theta_{\mathrm{fr}}(Y)}
{\Theta^{\displaystyle\prime}_{\mathrm{fr}}(Y)}
Q^{\displaystyle\prime}_{\mathrm{fr}}(Y).\label{52}
\end{align}
After ${\mathcal I}(Y)$ and ${\mathcal Q}(Y)$ are evaluated,
Eqs.~\eqref{parva} and~\eqref{49} give the exact voltage--charge relation $\widetilde V(Q)$ in parametric representation.
The asymptotic expansion becomes
\begin{equation}
\label{B21c}
{\mathcal I}(Y)\sim \sum_{p=1}^{\infty}
\frac{\bigl[-\Theta_{\mathrm{fr}}(Y)\bigr]^{p+1}}{(p+1)!}
\left[\frac{1}{\Theta_{\mathrm{fr}}^{\displaystyle\prime}(Y)}
\frac{d}{dY}\right]^pQ_{\mathrm{fr}}(Y).
\end{equation}

\section{Shock Profiles and the Incomplete Beta Function}
\label{ram}

\subsection{The Incomplete Beta Function}

The history of the incomplete beta function spans from 17th-century analysis to 20th-century statistical tables and modern computing.
Problems involving the quadrature of curves, analytically equivalent to the
evaluation of integrals of the form
\begin{equation}
\int_0^1 (1-x^p)^q\,dx
\end{equation}
(for integer $q$), were solved by John Wallis by 1655---before the invention
of the calculus by Newton and Leibniz. A study of Wallis' work led Newton to consider integrals of this type
in which the upper limit was replaced by a variable. The later history of the function is associated with the names of Gauss and Euler, while its modern formulation and systematic tabulation are largely associated with Karl Pearson~\cite{dutka}. Accordingly, below we consider the family of shock profiles for which
the voltage--charge relation and the squared local sound speed can be expressed in terms
of the lower incomplete beta function.

\subsection{The Asymmetric Power-Logistic Shock}

Let us consider a five-parameter family of profiles defined
in terms of the parameter $v\in(-1,1)$:
\begin{align}
\Theta&=\frac{(1+v)^{m_1}}{(1-v)^{m_2}},\label{f-profile}\\[2mm]
Q&=Q_0+G(1+v)^{n_1}+H(1-v)^{n_2}.\label{g-profile}
\end{align}
(In Sec.~\ref{ram} we assume $\Delta Q>0$.)
Here $Q_0$, $m_1>0$, $m_2>0$, $n_1>0$, and $n_2>0$
are five prescribed parameters.
The coefficients $G$ and $H$ are determined by the 
boundary conditions~\eqref{bbc}, and Eq.~\eqref{g-profile} takes the form
\begin{equation}
\label{general-profile}
\begin{split}
Q=Q_0&+2^{-n_1}(Q_2-Q_0)(1+v)^{n_1}\\[3mm]
&-2^{-n_2}(Q_0-Q_1)(1-v)^{n_2}.
\end{split}
\end{equation}
For $Q_0$ strictly between $Q_1$ and $Q_2$,
the function $Q_{\mathrm{fr}}(v)$ is strictly monotonic and
therefore invertible for $v\in(-1,1)$. In this generic case,
the asymptotic behavior near the edges is given by
Eqs.~\eqref{symptotic} and~\eqref{sympotic}, where
\begin{equation}
\label{uu}
\begin{cases}
\displaystyle \widetilde m_1=\frac{m_1}{\min\{n_1,1\}},\\[5mm]
\displaystyle \widetilde m_2=\frac{m_2}{\min\{n_2,1\}}.
\end{cases}
\end{equation}

Substituting Eqs.~\eqref{f-profile} and
\eqref{general-profile} into Eq.~\eqref{8h9} and~\eqref{52}, we obtain
\begin{align}
&{\mathcal I}(v)=2^{m_1-m_2}\Bigl[n_1(Q_2-Q_0)B_s(m_1+n_1,1-m_2)+\nonumber\\[4mm]
 &n_2(Q_0-Q_1) 
B_s(m_1+1,n_2-m_2)\Bigr],  \quad s=\frac{1+v}{2},\label{pardes}\\[5mm]
&{\mathcal Q}(v)=\frac{1-v^2}{m_1(1-v)+m_2(1+v)}\nonumber\\[4mm]
&\hskip 1cm\times\left[\frac{n_1(Q_2-Q_0)}{2^{n_1}(1+v)^{1-n_1}}
+\frac{n_2(Q_0-Q_1)}{2^{n_2}(1-v)^{1-n_2}}\right],   \label{pardes2}
\end{align}
where $B_s(a,b)$ is the lower incomplete beta function~\cite{temme},
\begin{equation}
B_s(a,b)\equiv\frac{1}{2^{a+b-1}}
\scaledint_{-1}^{v}(1+x)^{a-1}(1-x)^{b-1}\,dx.
\end{equation}

\subsection{The Two-Parameter Power-Logistic Shock}

Consider the two-parameter subfamily, characterized by the relations
\begin{equation}
\label{iu}
m_1=m_2=m, \qquad n_2=n, \qquad Q_0=Q_2.
\end{equation}
In this case, Eqs.~\eqref{f-profile} and~\eqref{general-profile} reduce to
the power-logistic profile
\begin{align}
\Theta&=\left(\dfrac{1+v}{1-v}\right)^m, \label{general}\\[2mm]
Q&=Q_2-2^{1-n}\Delta Q(1-v)^n,\label{general5}
\end{align}
introduced previously in Ref.~\cite{kogan6} and studied in Ref.~\cite{kogan7}.
In this case, the coefficient before $(1+v)^{n_1}$ in Eq.~\eqref{general-profile} vanishes, so Eq.~\eqref{uu} does not apply; directly from Eqs.~\eqref{general} and~\eqref{general5} one finds $\widetilde m_1=m$, $\widetilde m_2=m/n$. 

Substituting Eqs.~\eqref{pardes},~\eqref{pardes2} and~\eqref{general5} into Eq.~\eqref{48}, we obtain 
\begin{equation}
\label{wq2}
\begin{split}
&w=2^{n+1}mW\frac{(1-v)^{m-n}}{(1+v)^{m+1}}\\[2mm]
&\times B_s(m+1,n-m)+\frac{n-1+(n+1)v}{2m}.
\end{split}
\end{equation}
Equations~\eqref{general5} and~\eqref{wq2} give $w(Q)$ in the parametric representation.
Differentiating Eq.~\eqref{wq2} with respect to $v$, we obtain
\begin{equation}
\label{dir}
\begin{split}
w^{\displaystyle\prime}(v)=\frac{n+1}{2}\Biggl[4mW\frac{(1-v)^{m-n-1}}{(1+v)^{m+2}}\\[2mm]
\times\scaledint_{-1}^{v}\frac{(1+u)^m(v-u)}{(1-u)^{m-n+1}}\,du+\frac{1}{m}\Biggr]>0.
\end{split}
\end{equation}
Thus, the sound speed increases monotonically across the
power-logistic shock, from its value ahead of the shock
($v=-1$) to its value behind it ($v=1$). Consequently,
the positivity condition \eqref{soundd4} reduces to the condition \eqref{minus}
with $\widetilde m_1 = m$.

Using L'Hôpital's rule, we obtain:

\noindent
for $ 0<n<1$
\begin{equation}
\left.\frac{dw}{dQ}\right|_{Q=Q_2}=\left\{
\begin{array}{@{}l@{\quad}r@{}}
0, &\hspace{-5mm} m>2n,\\[3mm]
\dfrac{(n+1)mW}{2n\Delta Q}\,B(m+1,1-n),
   &\hspace{-5mm} m=2n,\\[5mm]
+\infty, &\hspace{-5mm} n<m<2n;
\end{array}
\right.
\end{equation}

\noindent
for $n=1$
\begin{equation}
\left.\frac{dw}{dQ}\right|_{Q=Q_2}=\left\{
\begin{array}{@{}l@{\quad}r@{}}
\dfrac{1}{\Delta Q}\left[\dfrac{mW}{(m-1)(m-2)}
+\dfrac{1}{m}\right],
   &\hspace{-5mm} m>2,\\[5mm]
+\infty, &\hspace{-5mm} 1<m\leq 2;
\end{array}
\right.
\end{equation}

\noindent
for $n>1$ and all $m>n$
\begin{equation}
\left.\frac{dw}{dQ}\right|_{Q=Q_2}=+\infty.
\end{equation}
Thus, if we require $\left.\frac{dw}{dQ}\right|_{Q=Q_2}$ to be finite, we should add to condition \eqref{plus} the restrictions: $n\leq1$, while $m\ge 2n$ for $n<1$ and $m>2$ for $n=1$.

\subsubsection{The hyperbolic-tangent shock} 
\label{hyperbola}

\begin{figure}[!t]
\centering
\includegraphics[width=\columnwidth]{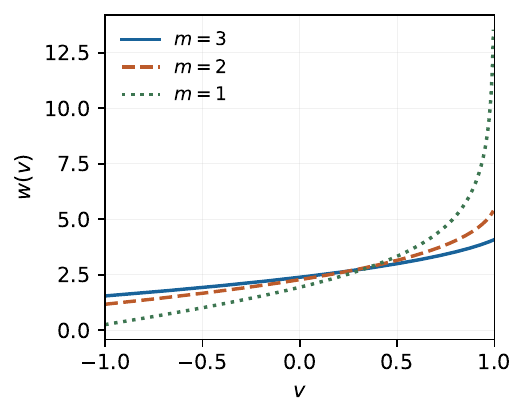}
\caption{The squared dimensionless local sound speed $w(v)$ defined by Equation~\eqref{eq-n1} for $m=1,2,3$ and
$W=5/2$. The solid, dashed, and dotted curves correspond to
$m=3$, $m=2$, and $m=1$, respectively; color is added only to improve
visual distinction.}
\label{fig:comparison}
\end{figure}

\begin{figure}[!t]
\centering
\includegraphics[width=\columnwidth]{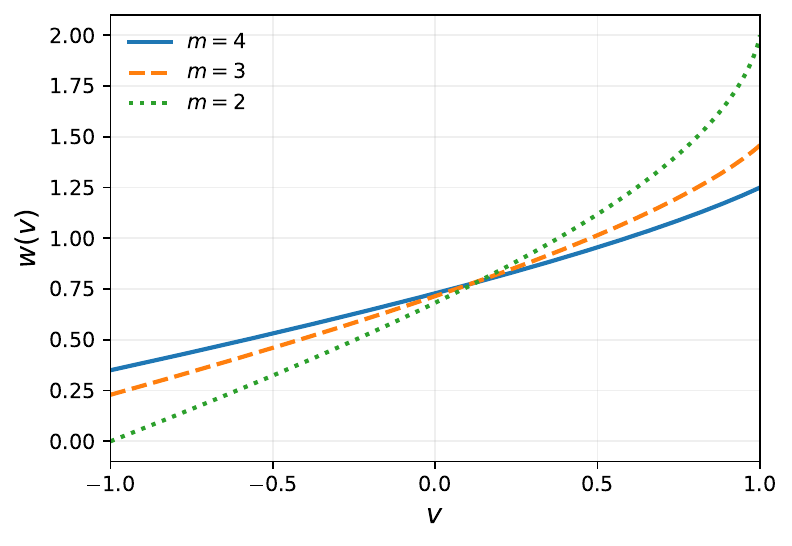}
\caption{The squared dimensionless local sound speed $w(v)$ defined by Equation~\eqref{eq-n1} for $m=2,3,4$ and
$W=3/4$. The solid, dashed, and dotted curves correspond to
$m=4$, $m=3$, and $m=2$, respectively; color is added only to improve
visual distinction.}
\label{fig:compar}
\end{figure}

We additionally restrict the subfamily considered in this subsection by assuming
$n=1$. 
Substituting $n=1$ into Eqs.~\eqref{general5} and~\eqref{wq2}, we obtain
\begin{align}
Q&=\overline Q+\Delta Q\,v,\label{hyperon}\\[2mm]
w&=4mW\frac{(1-v)^{m-1}}{(1+v)^{m+1}}
B_{\frac{1+v}{2}}(m+1,1-m)+\frac{v}{m}.\label{eq-n1}
\end{align}
In Figs.~\ref{fig:comparison} and~\ref{fig:compar}, we plot the squared dimensionless local sound speed $w(v)$. In Figure~\ref{fig:comparison} we see that for $m=1$, $w(Q)$ diverges when $Q\to Q_2$.
In Fig.~\ref{fig:compar} we see that $w(Q_1)=0$ for $m=2$ and $W=3/4$, in agreement with the condition $W=(m+1)/m^2$.

\section{The Direct Problem: Cubic Voltage-Charge Relation}
\label{direct-problem}

In this section, we consider the cubic voltage--charge relation
\begin{equation}
\label{cu}
\widetilde V(Q)=\widetilde V_c+\widetilde \alpha Q+\widetilde \beta Q^2+\widetilde \gamma Q^3,
\end{equation}
where $\widetilde V_c$, $\widetilde \alpha$, $\widetilde \beta$, and $\widetilde \gamma$ are constants. 
Here the additive constant of $\widetilde V$ is carried entirely by $\widetilde V_c$, i.e. we set $\widetilde V_0=0$ in Eqs.~\eqref{foy} and~\eqref{norma}.
We slightly generalize the boundary conditions~\eqref{bc} to
\begin{equation}
\label{bcc}
\bigl\{\lim_{\theta\to-\infty}Q(\theta),\;\lim_{\theta\to\infty}Q(\theta)\bigr\}
=\bigl\{Q_1,\;Q_2\bigr\}.
\end{equation}
Substituting the cubic relation~\eqref{cu} into Eq.~\eqref{199}
and introducing the dimensionless charge $v$ by Eq.~\eqref{hyperon},
we obtain
\begin{align}
v^{\displaystyle\prime\prime}(\theta)&
+\bigl[w\bigl(v(\theta)\bigr)+1\bigr]v^{\displaystyle\prime}(\theta)
=\mathcal P(v),\label{1992} \\[4mm]
w(v)&=\widetilde\alpha+2\widetilde\beta\,\overline Q
+3\widetilde\gamma\,\overline Q^2+2\Gamma v
+3\widetilde\gamma(\Delta Q)^2v^2,\label{77}\\[4mm]
\mathcal P(v)&\equiv C_0+C_1v -\Gamma\,v^2
-D\,v^3,\label{88b}
\end{align}
where
\begin{align}
\Gamma&\equiv(\widetilde\beta+3\widetilde\gamma\,\overline Q)\Delta Q,\\[3mm]
C_1&=W-\widetilde\alpha
-2\widetilde\beta\,\overline Q
-3\widetilde\gamma\,\overline Q^2,\\[2mm]
C_0&=\frac{W\overline Q-\widetilde V_c
-\widetilde\alpha\,\overline Q-\widetilde\beta\,\overline Q^2
-\widetilde\gamma\,\overline Q^3}{\Delta Q},\\[2mm]
D&\equiv\widetilde\gamma(\Delta Q)^2.
\end{align}
Substituting Eq.~\eqref{cu} into Eq.~\eqref{norma}, we obtain
\begin{equation}
\label{boundary}
\begin{aligned}
W&=\widetilde\alpha
+2\widetilde\beta\,\overline Q
+3\widetilde\gamma\,\overline Q^2+D,\\[3mm]
\widetilde V_c&=
\bigl(\widetilde\beta+2\widetilde\gamma\,\overline Q\bigr)
\bigl[\overline Q^2-(\Delta Q)^2\bigr].
\end{aligned}
\end{equation}
Using condition~\eqref{boundary}, Eq.~\eqref{77} becomes
\begin{equation}
\label{77-reduced}
w(v)=W+2\Gamma v+D(3v^2-1),
\end{equation}
and Eq.~\eqref{1992} takes the form
\begin{equation}
\label{1995}
v^{\displaystyle\prime\prime}(\theta)
+\bigl[w(v)+1\bigr]v^{\displaystyle\prime}(\theta)
=\bigl[\Gamma+Dv\bigr](1-v^2).
\end{equation}

\subsection{Weak-Shock Approximation}
\label{weak-shocks}

Assuming that $\Delta Q\to 0$, the solution of Eq.~\eqref{1995} in leading approximation is
given by Eq.~\eqref{197},
where in the considered case
\begin{equation}
m=m_{\mathrm ws}=\frac{W+1}{2\Gamma}\label{Gamma}.
\end{equation}
Since $W>0$, the sign of $m_{\mathrm{ws}}$ is the same as the sign
of $\Gamma$.
For $\Gamma>0$, the profile satisfies
$Q_{\mathrm{ws}}(-\infty)=Q_1$ and
$Q_{\mathrm{ws}}(+\infty)=Q_2$; for $\Gamma<0$, the orientation is
reversed. Thus, the sign of $\Gamma$ determines which asymptotic state
lies ahead of the front and which lies behind it. 

To check the obtained result, in the spirit of the paper, we use the trial profile~\eqref{197} in the inverse-design construction and calculate 
$w(v)$ which corresponds to that profile. More specifically, let us return to Sec.~\ref{hyperbola}
and consider broad hyperbolic-tangent shocks: $m\gg 1$. Substituting Eqs.~\eqref{general}
and~\eqref{hyperon}
into Eq.~\eqref{B21c}, and that into Eq.~\eqref{48}, we obtain
\begin{equation}
\label{64}
\begin{split}
w\sim W+\frac{(W+1)v}{m}+\frac{W}{2m^2}(3v^2-1)+\cdots.
\end{split}
\end{equation}
Substituting into Eq.~\eqref{64} the value of $m$ given by Eq.~\eqref{Gamma}, we obtain
\begin{equation}
\label{81b}
w\sim W+2\Gamma v+\frac{2W\Gamma^2}{(W+1)^2}(3v^2-1)+\dots.
\end{equation}
Comparing Eq.~\eqref{81b} with Eq.~\eqref{77-reduced} we started from,
we see that the first two terms in the equations agree. Under the compatibility condition
\begin{equation}
\label{compa}
D(W+1)^2 = 2W\Gamma^2,
\end{equation}
the third terms also agree. We return to Eq.~\eqref{compa} below.

In the two other approximations presented below, we use the weak-shock profile~\eqref{197} 
as a trial function, but consider $m$ as a free parameter.
Independently of the weak shock result, this trial function can be motivated as the simplest inverse profile with the required endpoint behavior. A monotone front connecting $v=-1$ and $v=1$ must have $dv/d\theta=0$ at both endpoints. Choosing the lowest-degree polynomial with these zeros gives $dv/d\theta=(1-v^2)/(2m)$, or equivalently $\theta_{\mathrm{tr}}^{\displaystyle\prime}(v)=2m/(1-v^2)$. 
The averaging approximation gives essentially the same result as the weak-shock approximation, but the variational approach improves the latter.

\subsection{Averaging Approximation}
\label{appro2}

Let us take the hyperbolic-tangent profile~\eqref{197} as an approximate solution of Eq.~\eqref{1995}, $m$ being treated as a fitting parameter. Substituting
\begin{equation}
v^{\prime}(\theta)=\frac{1-v^2}{2m}
\end{equation}
into Eq.~\eqref{1995} and dividing by $(1-v^2)$,
we obtain
\begin{equation}
\label{69}
\frac{W+1}{2m}-\Gamma+\left(\frac{\Gamma}{m}-\frac{1}{2m^2}
-D\right)v+\frac{D}{2m}(3v^2-1)=0.
\end{equation}
As a simple averaging approximation, we replace the factors $v$ and
$3v^2-1$ in this reduced condition by their averages over $(-1,1)$, both of
which vanish. We then obtain Eq.~\eqref{Gamma}.

If in Eq.~\eqref{69} only the term proportional
to $3v^2-1$ is replaced by its average, while the term linear
in $v$ is retained, matching the constant and linear terms yields
not only the value of $m$ given by Eq.~\eqref{Gamma}, but also the compatibility
condition among the prescribed parameters~\eqref{compa}.

\subsection{Variational Approximation}
\label{appro1}

In the framework of the variational approach, we start 
from Eq.~\eqref{1992}, without using Eq.~\eqref{boundary} a priori. 
The functional for minimization is
\begin{equation}
\label{acto-cubic}
{\mathcal F}[\theta^{\displaystyle\prime}(v)]=\frac{1}{2}\int\limits_{-1}^{1}
\mathcal R^2(v)\,dv,
\end{equation}
where 
\begin{align}
\mathcal R(v)&=\left[\frac{1}{\theta^{\displaystyle\prime}(v)}\right]^{\displaystyle\prime}
+w(v)+1-\mathcal P(v)\,\theta^{\displaystyle\prime}(v).\label{acto-cubic2} 
\end{align}
As the trial function, we will use the derivative of the power-logistic profile:
\begin{equation}
\label{cucu}
\theta_{\mathrm{tr}}^{\displaystyle\prime}(v)=\frac{2m}{1-v^2},
\end{equation}
considering $m$ as a fitting parameter.
This choice is motivated by the weak-shock approximation and by the fact that it
reduces the residual to a low-degree polynomial in $v$.

We need the integral
\begin{equation}
\int\limits_{-1}^{1}\left(\frac{\mathcal P(v)}{1-v^2}\right)^2\,dv
\end{equation}
to be finite.
Dividing $\mathcal P(v)$ by $(1-v^2)$, we obtain
\begin{equation}
\label{to-cubic}
\begin{split}
\mathcal P(v)=\left[\Gamma+D\,v\right](1-v^2)
+\left(C_0-\Gamma\right)+\left(C_1-D\right)v.
\end{split}
\end{equation}
Hence we need
\begin{equation}
\label{kondo}
\begin{aligned}
C_1-D&=0,\\[2mm]
C_0-\Gamma&=0.
\end{aligned}
\end{equation}
Equation~\eqref{kondo} coincides with Eq.~\eqref{boundary}; hence we recover Eq.~\eqref{77-reduced}.
Substituting the latter into Eq.~\eqref{acto-cubic2}, we obtain
\begin{align}
\mathcal R_{\mathrm{tr}}(v)=\bigl(W+1-2m\Gamma\bigr)
&+\left(2\Gamma-2Dm-\frac{1}{m}\right)v\nonumber\\[4mm]
&+D(3v^2-1).\label{81}
\end{align}
Substituting Eq.~\eqref{81} into Eq.~\eqref{acto-cubic} and
evaluating the integral, we obtain
\begin{equation}
\label{83}
\begin{split}
{\mathcal F}_{\mathrm tr}(m)=\left(W+1-2m\Gamma\right)^2
&+\frac{1}{3}\left(2\Gamma-2Dm-\dfrac{1}{m}\right)^2\\[2mm]
&+\frac{4}{5}D^2.
\end{split}
\end{equation}
The stationary values of $m$ are found from
\begin{equation}
\label{m-stationarity}
{\mathcal F}_{\mathrm tr}^{\displaystyle\prime}(m)=0.
\end{equation}
After differentiating and clearing the denominators, the stationarity condition becomes
\begin{equation}
\label{m-polynomial}
\begin{split}
&K(m)\equiv 4\left(3\Gamma^2+D^2\right)m^4\\[4mm]
&-2\Gamma\bigl[3(W+1)+2D\bigr]m^3
+2\Gamma m-1=0.
\end{split}
\end{equation}
Since $K(0)=-1<0$ and
$K(m)\to+\infty$ as $m\to\pm\infty$, the polynomial~\eqref{m-polynomial} has at least one positive and one negative real root, regardless of the sign of $\Gamma$. The other two
roots are either a complex-conjugate pair or both positive or both negative,
since the product of all four roots is negative. In any case, we choose 
the real root $m_{\mathrm{var}}$ that gives the smallest value of $\mathcal F_{\mathrm tr}(m)$. 

The solution of the equation
\begin{equation}
v^{\displaystyle\prime}(\theta)=\frac{1-v^2}{2m_{\mathrm{var}}},
\end{equation}
with the boundary conditions
\begin{equation}
\bigl \{\lim_{\theta\to-\infty}v(\theta),\;\lim_{\theta\to\infty}v(\theta)\bigr\}
=\bigl\{-1,\;1\bigr\}
\end{equation}
is obviously
\begin{equation}
v=\tanh\left(\frac{\theta}{2m_{\mathrm{var}}}\right).
\end{equation}

\subsection{Comparison Between the Approximations}

Under the compatibility condition~\eqref{compa},
all three approximations give the same result. To see this, we do not need to solve the algebraic equation~\eqref{m-polynomial}. Direct substitution shows that under the compatibility condition, $m_{\mathrm{ws}}$ is the global minimizer of ${\mathcal F}_{\mathrm tr}(m)$. Minimizing the sum of the two $m$-dependent terms on the right-hand side of Eq.~\eqref{83} and minimizing the first term alone give equivalent results,
because under the compatibility condition both terms reach absolute minimum for the same value of $m$. 

All three approximations also agree asymptotically as $\Delta Q\to 0$
for fixed
$\widetilde\alpha,\widetilde\beta,\widetilde\gamma$.
In this limit,
\begin{equation}
\Gamma=O(\Delta Q),\qquad D=O\bigl((\Delta Q)^2\bigr).
\end{equation}
Consequently, in the leading approximation, the polynomial
\eqref{m-polynomial} is
\begin{equation}
K(m)=12\Gamma^2m^4-6\Gamma(W+1)m^3,
\end{equation}
and the relevant root of the polynomial is given by Eq.~\eqref{Gamma}.
Under the compatibility condition, the mean-square residual after the minimization is of the order $O\left((\Delta Q)^4\right)$. 
In the general case, minimization gives the mean-square residual of order $O\left((\Delta Q)^2\right)$. 
So, under the compatibility condition, the minimizing solution
is parametrically more precise than in the general case. 

The approximations give different results when the compatibility condition does not hold.
The variational approximation directly minimizes the specified squared-residual functional. It is also less ad hoc and can be improved by using a trial function with more than one fitting parameter. We leave a systematic study of the variational approximation for future work.

\section{Spectral Problem}
\label{dicho}

\subsection{Local Spectral Analysis}

In this section, we continue to study the direct problem. 
We consider small perturbations of the traveling-front solution 
$(Q_{\mathrm{fr}},U\widetilde I_{\mathrm{fr}})$ of Eq.~\eqref{c2b},
discarding any previously imposed restrictions on the latter or on the function 
$\widetilde V(Q)$.
In the frame moving with the front we write the perturbed solution as
\begin{equation}
\label{absolute}
\begin{aligned}
&Q(z,t)=Q_{\mathrm{fr}}(\theta)+\epsilon\,\phi(\theta)\exp\left(
\frac{\lambda R_Lt}{L}\right),\\[2mm]
&I(z,t)=U\left[\widetilde I_{\mathrm{fr}}(\theta)
+\epsilon\,\psi(\theta)\exp\left(\frac{\lambda R_Lt}{L}\right)\right],
\end{aligned}
\end{equation}
where $|\epsilon|\ll1$.
Substituting Eq.~(\ref{absolute}) into Eq.~(\ref{c2b}) and linearizing, we obtain
a variable-coefficient eigenvalue problem:
\begin{equation}
\label{oud}
\begin{aligned}
&\left[\psi^{\displaystyle\prime}(\theta)
+\psi(\theta)-\phi(\theta)\right]^{\displaystyle\prime}
=\lambda\left[\phi(\theta)
-\psi^{\displaystyle\prime}(\theta)\right],\\[4mm]
&\Bigl[\phi^{\displaystyle\prime}(\theta)
+w\bigl[Q_{\mathrm{fr}}(\theta)\bigr]\phi(\theta)
-W\psi(\theta)\Bigr]^{\displaystyle\prime}\\[2mm]
&\hskip 3.2cm =\lambda\left[W\psi(\theta)
-\phi^{\displaystyle\prime}(\theta)\right].
\end{aligned}
\end{equation}
For the point spectrum, we seek localized eigenfunctions satisfying
the boundary conditions 
\begin{equation}
\begin{cases}\phi(\theta)\longrightarrow 0,\\[2mm]
\psi(\theta)\longrightarrow 0,
\end{cases}\qquad\theta\longrightarrow\pm\infty.
\end{equation} 
Translational invariance implies the neutral
eigenvalue $\lambda=0$, with eigenfunctions
\begin{equation}
\begin{cases}
\phi(\theta)=Q_{\mathrm{fr}}^{\displaystyle\prime}(\theta),\\[3mm]
\psi(\theta)=\widetilde I_{\mathrm{fr}}^{\displaystyle\prime}(\theta),
\end{cases}
\end{equation}
which becomes obvious when we compare Eq.~(\ref{oud}) with Eq.~(\ref{low}). 
Spectral stability in the moving frame requires that, apart from
the translational eigenvalue at $\lambda=0$, the spectrum contains
no points with $\operatorname{Re}\lambda>0$.
The local analysis below does not constitute a proof that the full
variable-coefficient problem has no unstable global point spectrum.

We now examine the local short-wavelength behavior of Eq.~(\ref{oud}) using a frozen-coefficient approximation (equivalently, the leading-order WKB approximation) appropriate to wavelengths short compared with the front
width. Specifically, we will treat $w\bigl(Q_{\mathrm{fr}}(\theta)\bigr)$ as locally constant. Its dependence
on $\theta$ then enters only parametrically through its local value.

We seek normal-mode solutions of the form
\begin{equation}
\label{ali}
\begin{split}
\phi(\theta)=\widehat{\phi}\,\exp(ik\theta),\\[2mm]
\psi(\theta)=\widehat{\psi}\,\exp(ik\theta),
\end{split}
\end{equation}
where $k\in\mathbb R$, and $\widehat{\phi}$ and $\widehat{\psi}$ are constants.
The resulting dispersion relation is
\begin{equation}
\label{beb}
W(\lambda+ik)^2
+k^2(1+\lambda+ik)\left[w+\lambda+ik\right]=0,
\end{equation}
where we have suppressed the argument $Q_{\mathrm{fr}}(\theta)$ of $w$.
Hence, the two spectral modes are
\begin{equation}
\label{beba}
\lambda_\pm=-ik+\frac{-k^2(1+w)
\pm\sqrt{k^4(1-w)^2-4Wwk^2}}
{2(W+k^2)}.
\end{equation}
For $W>0$ and $w>0$, both roots of the
local frozen-coefficient dispersion relation have negative real parts.
Thus all nonzero modes described within the range of validity of the
frozen-coefficient approximation are damped.

\subsection{The Kink--Shock Dichotomy}
\label{stab}

In either asymptotic state, where $Q_{\mathrm{fr}}=Q_i$,
$w$ is constant. Hence Eqs.~(\ref{beb}) and~(\ref{beba}) in these states
are exact and valid for arbitrary real $k$, without invoking the
short-wavelength approximation. We now consider long-wavelength perturbation modes in these states. Expanding Eq.~(\ref{beba}) as $k\to0^+$
and retaining the leading term, we obtain
\begin{equation}
\lambda_{\pm,i}=-ik\pm ik\sqrt{\frac{w_i}{W}}+O(k^2)=ik\left(-1\pm\frac{u_i}{U}\right)+O(k^2),
\end{equation}
where $u_i\equiv u(Q_i)$.
For fixed $t$, the complete perturbation factor in the long-wavelength limit
$k\to0^+$ can be written as
\begin{equation}
\begin{aligned}
\exp\left(ik\theta+\frac{R_L}{L}\lambda_{\pm,i}t\right)
\sim\exp\left[i\frac{kR_L}{LU}
\Bigl((\pm u_i-U)t-\widetilde z\Bigr)\right],
\end{aligned}
\end{equation}
where $\widetilde z=z-Ut$.
Thus, the velocities of the two characteristic waves relative to the front are $\pm u_i-U$.

Focusing on the mode that propagates in the laboratory frame in the same direction as the front makes the distinction between kinks and shocks transparent. In the case
\begin{equation}
u_1<U<u_2,
\end{equation}
the front overtakes the long-wavelength disturbances in the state
ahead of it, whereas such disturbances in the state 
behind it overtake the shock. Thus, characteristics approach the shock from both asymptotic states, and the shock is compressive~\cite{courant}.
By contrast, for 
\begin{equation}
U>\max\{u_1,u_2\},
\end{equation}
the front overtakes disturbances in both
asymptotic states and is therefore noncompressive.

\section{Conclusions}

We formulated a traveling-wave equation describing traveling fronts in
a lossy nonlinear transmission line with a nonlinear voltage--charge
relation. Starting from the continuum Kirchhoff equations, we reduced
the traveling-wave problem to a second-order Li\'enard equation for the
charge profile. This equation provides a mechanical interpretation of
the front dynamics and forms the basis for the variational treatment of
the direct problem.

In the inverse problem, the front profile and its dimensionless squared
speed $W$ were prescribed, while the voltage--charge relation
$\widetilde V(Q)$ was determined as part of the solution. We obtained
exact expressions for both $\widetilde V(Q)$ and the dimensionless
squared local sound speed $w(Q)$, whose finiteness and positivity
determine the physical admissibility of the front. These expressions apply to any
sufficiently smooth, monotone, and invertible front profile.

We studied in detail a five-parameter asymmetric power-logistic family.
For this family, the voltage--charge relation and the squared local sound speed
can be expressed in terms of the lower incomplete beta function. The
corresponding admissibility conditions show that the physically
admissible fronts in this family are shocks rather than kinks. The
previously studied symmetric hyperbolic-tangent profile is recovered as
a particular case.

For the direct problem, in which the voltage--charge relation is
prescribed, we formulated a variational method for approximately solving
the traveling-wave equation by minimizing its squared residual. 

Finally, we derived the variable-coefficient spectral problem by
linearizing the original partial differential equations about the
traveling front. The frozen-coefficient dispersion relation shows that
the local short-wavelength modes are damped when $W>0$ and $w(Q)>0$.
In the asymptotic states, the long-wavelength characteristic velocities
reproduce the distinction between compressive shocks and
noncompressive kinks obtained from the mechanical analogy. The local
analysis does not, however, exclude unstable global point-spectrum
modes; establishing full spectral stability remains an important
problem for future work.

\section*{Acknowledgments}
The author is grateful to S.~Flach, M.~Goldstein, P.~G.~Kevrekidis and B. Malomed for helpful discussions.

\end{document}